# Observation of Young's double-slit phenomenon in anti-PT-symmetric electrical circuits

Keyu Pan,[1] Xiumei Wang,[2*] Xizhou Shen,[1] Haoyi Zhou,[2] and Xingping Zhou[3*]

[1] *College of Integrated Circuit Science and Engineering, Nanjing University of Posts and Telecommunications, Nanjing 210003, China*

[2] *College of Electronic and Optical Engineering, Nanjing University of Posts and Telecommunications, Nanjing 210003, China*

[3] *Institute of Quantum Information and Technology, Nanjing University of Posts and Telecommunications, Nanjing 210003, China.*

*wxm@njupt.edu.cn

*zxp@njupt.edu.cn

In the last few decades, interference has been extensively studied in both the quantum and classical fields, which reveals light volatility and is widely used for high-precision measurements. We have put forward the phenomenon in which the discrete diffraction and interference phenomena, presented by the time-varying voltage of a Su-Schrieffer-Heeger (SSH) circuit model with an anti-PT (APT) symmetry. To demonstrate Young's double-slit phenomenon in an APT circuit, we initially explore the coupled mode theory (CMT) of voltage in the broken phase, observe discrete diffraction under single excitation and interference under double excitations. Furthermore, we design a phase-shifting circuit to observe the effects of phase difference and distance on discrete interference. Our work combines the effects in optics with condensed matter physics, show the Young's double-slit phenomenon in electrical circuits theoretically and experimentally.

Interference is an important concept in wave theory, occurring when two or more waves overlap in space and combine to form new waveform phenomena. In 1807, British scientist Thomas Young proposed that when light passes through two narrow slits, the resulting light waves will overlap and interfere, forming interference fringes with alternating brightness and darkness [1]. This discovery revealed the wave nature of light and laid the foundation for wave theories. With the emergence of quantum mechanics, the wave-particle duality of light has been proven, expanding the research platform for double-slit interference to electron beams [2-4], photons [5-8], surface plasmon polaritons (SPP) [9-10], etc.

Parity-Time (PT) symmetry is a common phenomenon in non-Hermitian systems, where the balance between gain and loss leads to real-valued characteristic energies in symmetric phases [11]. On the other hand, anti-PT (APT) symmetry, as a concept opposite to PT symmetry, represents another intriguing class of non-Hermitian systems [12-19]. Such systems exhibit purely imaginary eigenenergies in symmetric phases and give rise to level attraction [20], constant refraction [21], energy-conserving difference dynamics [22], and chiral mode switching for symmetry-broken modes [23]. From the perspective of lattices, the Su-Schrieffer-Heeger (SSH) model [24] is considered as the simplest model in topological state research [17, 25-29]. Each unit cell consists of two lattice sites, and the hopping term connects two different sublattices.

In our work we have put forward a phenomenon in which the discrete diffraction and interference phenomena, presented by a SSH circuit model with APT symmetry. The utilization of the negative impedance converter with current inversion (INIC) and resistors facilitates the straightforward incorporation of effective gain and loss in the circuit, thereby enabling non-Hermitian coupling [30-37]. Meanwhile, the topological circuit systems offer flexibility in designing various physical systems. Simultaneously, all parameters can be precisely tuned by taking advantage of modern technology in integrated circuits [38]. Many phenomena such as Bloch oscillations and dipole oscillations in optics [39, 40], exceptional points in condensed matter physics [41], non-Hermitian skin effect [42], Anderson localization [43] and two orbitals system [44] have been implemented in the circuit. In order to show Young's double-slit phenomenon in the circuit, we initially investigate the coupled mode theory (CMT) of system voltage based on Kirchhoff's law and Bloch's theorem. We select the APT circuit [45, 46] in the broken phase to realize Young's double-slit interference. Firstly, we observe discrete diffraction phenomena under the single-node excitation. Then, we explore the interference phenomena under two excitations. Furthermore, we design a phase-shifting circuit to investigate the influence of phase difference between two excitations and their distance on the discrete interference phenomena.

The proposed circuit consists of a series of RLC parallel resonators, one end connected to the ground, the other to the adjacent resonators through the resistance coupling, which converts electrical energy into thermal energy, acting as a non-Hermitian coupling. We set two nodes as one unit, where all the resonators connected by odd nodes have the same electrical parameters, and all the resonators connected by even nodes have another set of identical electrical parameters. For the *n*th group of nodes, we set the voltages at odd and even nodes, by $V_1^n$ and $V_2^n$ respectively. We set up an odd number 377 of nodes, and the two ends of the circuit chain are coupled to the ground by resistance. The detailed structure of the circuit system is shown in Fig. 1(a). It is important to note that the resistance of the RLC resonator is a negative group realized by an INIC, consists of an operational amplifier accompanied by a negative feedback network, as sketched in Fig. 1(b). Unlike the energy consumed by a positive resistor, the negative resistor provides a gain that injects energy into the circuit.

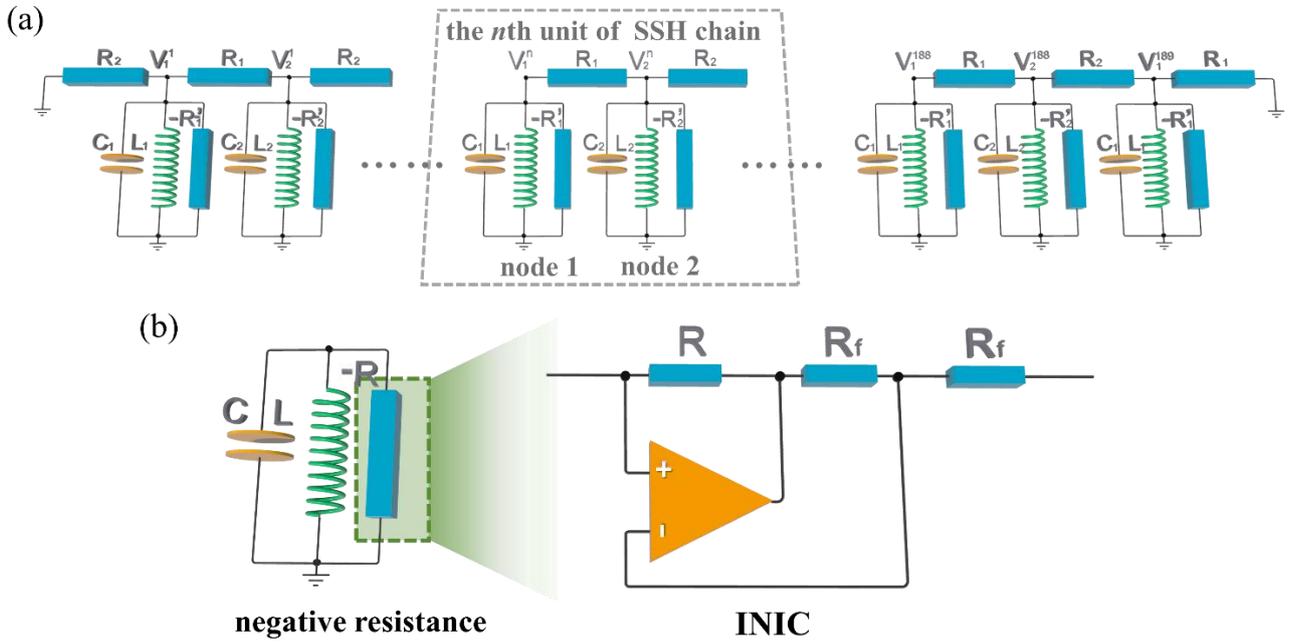

Fig. 1. (a) A SSH circuit chain consisting of a coupled RLC resonator with an APT symmetry. The resonators are coupled by resistance, and the resistance within the resonator is negative to provide energy gain. (b) A negative resistance implemented through the INIC, where the INIC is composed of an operational amplifier and a negative feedback network.

Based on Kirchhoff's law, the temporal evolution of voltage at any node can be obtained.

$$\begin{aligned}\frac{d^2 V_1^n}{dt^2}+\omega_1^2 V_1^n &= \frac{1}{R_2 C}\frac{dV_2^{n-1}}{dt}+\frac{1}{R_1 C}\frac{dV_2^n}{dt} \\ \frac{d^2 V_2^n}{dt^2}+\omega_2^2 V_2^n &= \frac{1}{R_1 C}\frac{dV_1^n}{dt}+\frac{1}{R_2 C}\frac{dV_1^{n+1}}{dt}\end{aligned} \quad (1)$$

where $\omega_1 = 1/\sqrt{L_1 C}$ and $\omega_2 = 1/\sqrt{L_2 C}$ denote the resonant of the RLC circuits. Based on the Bloch's theorem, we get the periodic condition $V_1^n = V_1^{n-1} e^{i\varphi}$ and $V_2^n = V_2^{n-1} e^{i\varphi}$. Under the condition of the weak coupling $1/R_{1,2}C \ll 1/\sqrt{L_{1,2}C}$, the small-detuning regimes $|\omega_1 - \omega_2| \ll |\omega_1 + \omega_2|$ and ignoring the second-order partial guide terms in the equation, we can obtain as follows

$$i\frac{d}{dt}\begin{pmatrix} V_1^n \\ V_2^n \end{pmatrix} = \begin{pmatrix} \dfrac{\omega_1 - \omega_2}{2} & \dfrac{i}{2R_1 C} + \dfrac{i}{2R_2 C} e^{-i\varphi} \\ \dfrac{i}{2R_1 C} + \dfrac{i}{2R_2 C} e^{i\varphi} & \dfrac{\omega_2 - \omega_1}{2} \end{pmatrix} \begin{pmatrix} V_1^n \\ V_2^n \end{pmatrix}. \tag{2}$$

Thus, the proposed structure is in the form of APT. We can obtain the phase diagram by calculating the eigenvalues of the system, which can be written as:

$$\omega = \pm\frac{1}{2}\sqrt{(\omega_1 - \omega_2)^2 - (R_1 C)^2 - (R_2 C)^2 - 2R_1 R_2 C^2 \cos\varphi}. \tag{3}$$

The positive and negative in the root number determine the virtual part of the eigenvalue. According to the different circuit parameters, the phase diagram is divided into three parts, respectively corresponding to the eigenvalue spectrum is the pure virtual part (Broken Area), and the virtual part exists simultaneously (Intermediate Area) and the pure real part (Symmetric Area) [45, 46].

Then, the band structures have an imaginary part in symmetric and intermediate phases, which leads to an exponential growth of the total power. In the broken phase, the band structures are purely real-valued, the total power is stable to some extent and exhibits oscillations due to the non-orthogonal eigenstates [45, 46]. The voltage time evolves rapidly and surges at an exponential rate in a very short time. Only a discrete diffraction can be observed in the broken phase. We set the number of resonators to 377. The parameters of the circuit system are set as $R_1=R_2=1000$ Ω, $C_1=C_2=C=470$ nF, $L_1=1.5$ mH and $L_2=1$ mH, to fulfil the condition in the broken phase and slowing envelops. We take the MATLAB Simulink for the circuit simulation. We set the initial voltage of the middle capacitor (resonator number 189, or node voltage count as) to 1V. After 0.4 ms, we record the time-varying voltage of each node in the whole circuit lattice chain, as shown in Fig. 2(a). To confirm this result, we also perform theoretical calculations based on CMT. The time-varying voltage of the circuits lattice chain can be described by the following set of partial differential equations in the tight-binding approximation (TBA) [47]:

$$\begin{cases} \dfrac{dV_1^n}{dt} = \dfrac{\omega_2 - \omega_1}{2}V_1^n + \dfrac{1}{2R_1C}V_2^n + \dfrac{1}{2R_1C}V_2^{n-1} \quad (n=2,3,...188) \\ \dfrac{dV_2^n}{dt} = \dfrac{\omega_1 - \omega_2}{2}V_2^n + \dfrac{1}{2R_1C}V_1^n + \dfrac{1}{2R_1C}V_1^{n+1} \quad (n=1,2,...188) \end{cases} \quad (4)$$

$$\dfrac{dV_1^1}{dt} = \dfrac{\omega_2 - \omega_1}{2}V_1^1 + \dfrac{1}{2R_1C}V_2^1, \quad \dfrac{dV_1^{189}}{dt} = \dfrac{\omega_1 - \omega_2}{2}V_1^{189} + \dfrac{1}{2R_1C}V_2^{188}.$$

We set the initial conditions $V_1^{95} = 1$. The results are shown in Fig. 2(d), which is completely consistent with the analytical calculation results based on MATLAB Simulink in Fig. 2(a). Starting from a single excitation capacitance, the power is transferred to the adjacent resonator via the non-Hermitian coupling of the circuit. These resonators are coupled to their neighbors in turn by a single resistance. We observe the phenomenon of discrete diffraction, which is a characteristic pattern occurred only in discrete systems.

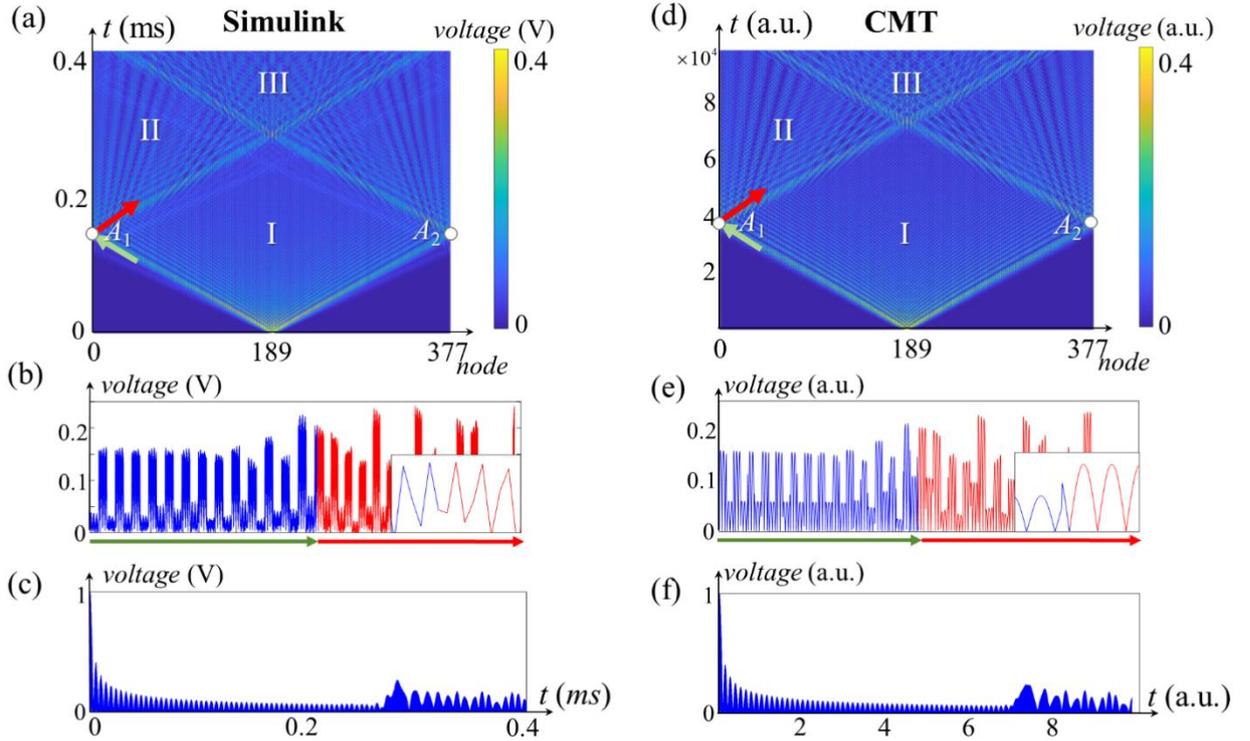

Fig. 2. (a) Discrete diffraction and interference phenomena of the time-varying voltage evolution under a single excitation. (b) Voltage changes before and after the reflection point. (c) Temporal voltage changes at the central node. (d-f) The results of theoretical calculations based on CMT respectively correspond to (a-c)

When the time duration is sufficiently long, we can observe more than discrete diffraction solely in region I in Fig. 2(a). As our constructed circuit system has finite length, the energy undergoes reflection upon reaching the boundaries at both ends of the circuit indicated by Point $A_1$ and $A_2$ in Fig. 2(a). To investigate the phenomenon of half-wave loss at the reflection point, we scan along the coloured line representing the propagation of initial energy in the resonator before and after reflection, as shown in Fig. 2(a-b). The voltages on the diagonal lines before and after reflection respectively correspond to the blue and red curves in Fig. 2(b). The voltage variation near the reflection point is depicted in the inset. It can be observed that at the junction of these two curves, there is a smooth voltage change without a sudden phase shift of 180 degrees. Thus, half-wave loss does not occur during the reflection.

Furthermore, interference fringes are formed in region II, which can be attributed to the reflection occurring at the edges. It creates an initial voltage at the external nodes formed by the voltage of the central node and there exists an equivalent mirror excitation. More complex multislit interference occurs in region III. In Fig. 2(c), we present the time evolution of voltage at the central node (node 189). There is a rapid decay of instantaneous excitation followed by oscillations within the diffraction region and subsequently an evident increase in voltage within the interference region. The results of theoretical calculations based on CMT are respectively depicted in Fig. 2(d-f), which are consistent with the circuit simulations.

There is not only discrete diffraction but also interference in the APT symmetric circuit. Then we specifically explore the discrete interference. We choose the same parameters used in Fig. 2. We apply an initial voltage of 1V to the two capacitors at symmetric nodes (node 151 and node 227). The MATLAB Simulink results of the circuit are shown in Fig. 3(a). Under two symmetric excitations, the time-varying voltage of each node exhibits distinct equidistant bright and dark fringes, indicating clear discrete interference phenomena. Due to the presence of higher harmonics, the intensity of each fringe varies in the circuit system's generated time-varying voltage patterns, similar to mixed light interference in optics with multiple frequencies. The theoretical calculation results based on CMT are also presented in Fig. 3(b), which align well with the Simulink results.

In addition, we extract the voltage of each node at a certain moment and plot the curve of node number versus voltage, as depicted in the inset of Fig. 3(a-b). It can be observed that the variation in voltage follows a pattern consistent with optical interference intensity, with distinct dark and bright fringes evenly spaced within the interference region.

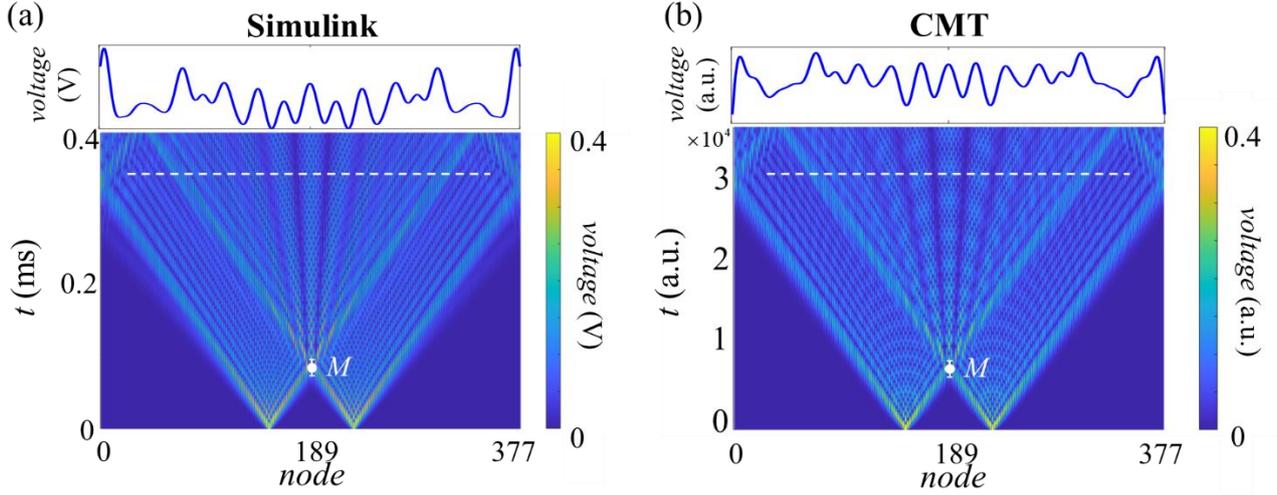

Fig. 3. (a) MATLAB Simulink calculated the time-varying voltage with symmetrically double excitation. (b) CMT calculated the time-varying voltage with symmetrically double excitation.

Next, we conduct a qualitative analysis to verify whether it satisfies the optical interference characteristics. We assume two initial voltage excitations as $v_m$ and $v_n$ with $v_m = v_n = v_0$, and the distance between the two excitations is denoted as $d$, which corresponds to the number of nodes between them. We denote the time as $D$, which is equivalent to the distance from the light source to the receiving screen in optical double-slit interference [1]. We can obtain the following relationship:

$$V = v_m + v_n + 2\sqrt{v_m v_n} \cos \Delta\varphi$$
$$\Delta\varphi = k(r_1 - r_2) = k\delta = \frac{2\pi}{\lambda}\delta \quad (5)$$

where $V$ is the voltage at each point, $\Delta\varphi$ is the phase difference. The $r_1, r_2$ represent the pathway distance of energy propagation and the pathway distance difference $\delta$ can be obtained by the following formula:

$$r_1 = \sqrt{D^2 + \left(x - \frac{d}{2}\right)^2}, \quad r_2 = \sqrt{D^2 + \left(x + \frac{d}{2}\right)^2}, \quad (6)$$

$$\delta = r_2 - r_1 = \frac{r_2^2 - r_1^2}{r_1 + r_2} = \frac{2xd}{\sqrt{D^2 + \left(x - \frac{d}{2}\right)^2} + \sqrt{D^2 + \left(x + \frac{d}{2}\right)^2}}. \quad (7)$$

When $D \gg d$, we can approximately obtain $\delta \approx d\sin\theta$. If the $\theta$ is small enough, we approximately obtain $\sin\theta \approx \tan\theta = \frac{x}{D}$, and further get $\delta = r_2 - r_1 \approx d\sin\theta \approx \frac{xd}{D}$. Then, we have the voltage at any time at any node

$$V = 4V_0 \cos^2(\frac{\pi}{\lambda}\frac{xd}{D}). \tag{8}$$

When $\Delta\varphi = 2m\pi$, equally $\delta = m\lambda$ $(m = 0, \pm1, ...)$, we obtain $V = 4V_0$ corresponding to the strengthening point, showing a bright stripe. When $\Delta\varphi = (2m-1)\pi$, equally $\delta = \left(m - \frac{1}{2}\right)\lambda$ $(m = \pm1, \pm2...)$, we obtain $V = 0$ corresponding to the weakening point, showing a dark stripe.

To see more clearly, we select a specific time (indicated by the white dashed line in the Fig. 3) with $d = 76$, $D = 0.33$ ms, $\lambda = \frac{c}{f_0} = 5.12 \times 10^4$ m, where $f_0$ is the fundamental frequency. We obtain $\Delta x = 25$, which is consistent with the simulation results depicted in the inset of Fig. 3(a-b), based on the parameters mentioned above and combined with formula $\Delta x = x_{m+1} - x_m = \frac{(m+1)\lambda D}{d} - \frac{m\lambda D}{d} = \frac{D\lambda}{d}$.

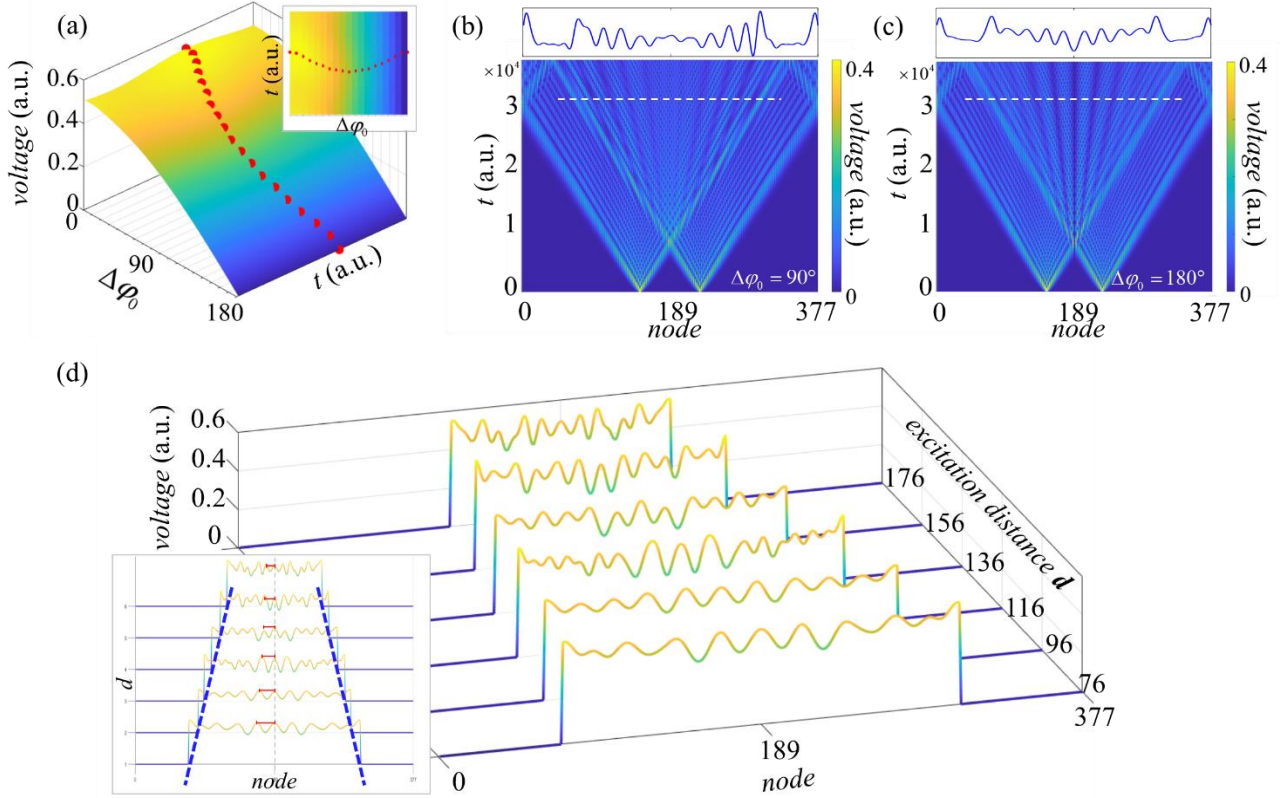

Fig. 4. Effects of phase difference and distance between two excitations on interference by the results of CMT. (a) Influence of different phase differences between two excitations on the voltage near the interference point. (b-c)

Voltage distribution for excitation phase differences of 90 and 180 degrees, respectively. (d) Influence of different distances between two excitations on the voltage at various nodes at a certain moment.

Furthermore, we make changes to certain variables and observe their effects on interference. Firstly, we alter the phase difference between the two excitations with a phase-shifting circuit, which keeps the signal amplitude remains unchanged, and the phase realizes $0°$ to $180°$ adjustable lag. The specific circuit structure and working principle are shown in Appendix A. We simulate the circuit model with 19 sets of different phase shift angle with a step size of $10°$, ranging from 0 to $180°$ through MATLAB Simulink and CMT theory respectively. In order to explore the influence of the phase difference on discrete interference, we take some points at the moment near the point $M$ indicated in Fig. 3(a) and draw the relationship between the phase difference and the corresponding voltage shown in Fig. 4(a). We observe that the interference at the central node gradually changes from enhancement to weakening, and the voltage keeps decreasing, as the phase difference increases. When the excitation has an initial phase difference $\Delta\varphi_0$, we get $\Delta\varphi = \frac{2\pi\delta}{\lambda} + \Delta\varphi_0 = \frac{2\pi}{\lambda}\frac{xd}{D} + \Delta\varphi_0$. The position of the clear stripes has changed from $x = m\frac{D\lambda}{d}$ to $x' = m\frac{D\lambda}{d} - \frac{\Delta\varphi_0 \cdot D\lambda}{2\pi d}$. Originally, the time-varying voltage exhibits a symmetric distribution with $\Delta\varphi_0 = 0$ shown in Fig .3(b). As the $\Delta\varphi_0$ increases, the stripes shift and the time-varying voltage is no longer symmetrically distributed. For example, the phenomenon at $\Delta\varphi_0 = 90°$ of asymmetric distribution of stripes is illustrated in Fig .4(b). At $\Delta\varphi_0 = 180°$, the dark stripes precisely shift to the center node position, restoring a symmetric distribution as shown in Fig. 4(c).

Next, we adjust the distance between the two excitations (keeping the phase difference at 0) and observe its impact on interference. We select seven sets of data with excitations distance of 76, 96, 116, 136, 156, and 176. We show the curves of node-voltage at the same time indicted by dashed white line in Fig. 5(b-c) and plot a waterfall chart in Fig. 5(d) (setting non-interference areas to a value of 0 for better visibility). The red line in the inset of Fig. 5(d) reflects the phenomenon that when the excitations distance $d$ increases, the stripe spacing decreases continuously, due to the theoretical calculation $\Delta x = \frac{D\lambda}{d}$. The small acceptable error comes from the dispersion of the circuit nodes. The blue dashed line represents that as the excitations distance $d$ increases, the interference occurs later in the time domain.

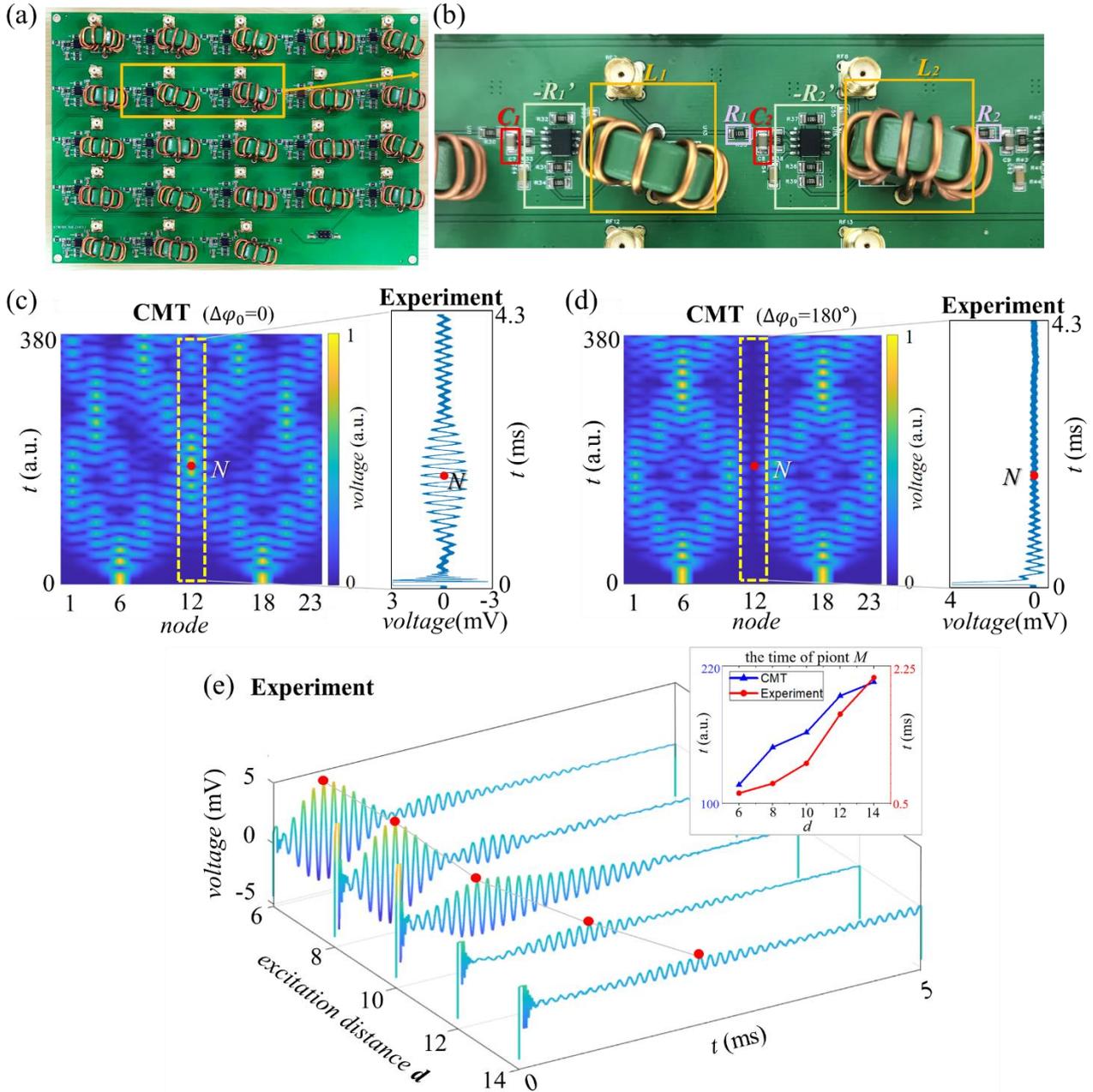

Fig. 5 The experimental realization of the circuit chain and the interference phenomenon presented by the node energy evolution on PCB. (a) The photograph of the PCB fabricated with 23 resonators of APT symmetry. (b) A magnified view of one unit depicting the circuits and components of the system. (c) Discrete interference enhancement phenomenon of voltage evolution, with the left is results of CMT and the right is the experimental results of the central node 12. (d) The destructive interference phenomenon with 180° phase difference of the two injections. The left is results of CMT and the right is the experimental results of the center node 12. (e) Voltage evolution of the central node under 5 sets of different distance injections.

In order to verify the discrete interference phenomenon in the circuit, we construct a SSH circuit with 23 RLC resonators in series, and make a printed circuit board (PCB), shown in Fig. 5(a). Detailed structure of one unit (consisting of two RLC resonators) is annotated in Fig. 5(b) to identify the capacitors (red boxes), the inductors (orange boxes), the positive resistance used for the coupling (pink boxes) and the INICs (light green boxes). The circuit parameters are set to $R_1=R_2=1000$ Ω, $C_1=C_2=C=470$ nF, $L_1=1.35$ mH and $L_2=1$ mH. The INIC within our experimental setup is composed of TL081ACD operational amplifiers, provided by Texas Instruments. It is worth noting that the inductance here is customized, which has a large inductance while the internal resistance is less than 0.02 Ω. If the internal resistance is too large, loss will be introduced, resulting in rapid attenuation of the node voltage, and interference phenomenon cannot be observed. For the setting of the incentives, we directly set the initial voltage of the capacitor in the circuit simulation, while in the experiment, we inject the pulse current into the two symmetric positions (node 6 and node 18), so as to quickly charge the capacitor and form the initial voltage. Note that pulsed voltage sources cannot be used, as a pulse followed by a voltage of 0 will cause the injected node voltage to remain 0. At the same time, due to the limited output power of the signal generator, the observed voltage will decay rapidly, leading the interference phenomenon cannot be observed. Therefore, we use the voltage controlled constant current source made by the OPA549 power amplifier to amplify the output pulse signal of the signal generator. The two power respectively supply energy to the constant current source module and the INIC module of the test PCB.

In the experiment, we focus on measuring the voltage of the central node (node12), and observe the phenomenon of discrete interference through its voltage evolution. We inject pulse current at node 6 and node 18 without any phase difference in Fig. 5(c), the experimental results are shown on the right, which is highly similar to the boxed and highlighted part of the results of CMT in the left figure. It can be seen the voltage increases significantly after a period of time, and reaches its maximum at point $N$, which is a phenomenon of interference enhancement. At the same time, we also measure the voltage of the intermediate node at the 180° phase difference of the two injections. The CMT and experimental measurement results are shown in Fig.5 (d), which exhibit the destructive interference. Excluding the interference of clutter, the voltage at point $N$ is almost 0. Finally, we vary the injection distance and measure the central node voltage under 5 sets of different distance injections shown in Fig. 5(e) as well as the time of the point $N$ in Fig. 5(c), which is marked with red dots and specially depicted in the inset of Fig. 5(e). The error in the experiment may come from many aspects. The measured node voltages may exhibit slight discrepancies compared to the simulation data, which can be attributed to the inherent tolerance in the circuit components—inductors with a 20% margin of error and capacitors with a 10% margin, as well as the precision limits of the measurement apparatus. Besides we use pulse current injection in the experiment instead of setting initial voltage of the simulation, which will also bring small errors in time. However, despite these variances, the overall experimental outcomes align with the theoretical model, indicating a successful observation of discrete double-slit interference phenomenon within our meticulously designed circuit. Additional details about the specific measurement instruments can be found in Appendix B.

In conclusion, we have put forward a phenomenon in which the discrete diffraction and interference phenomena, presented by the time-varying voltage of the nodes, can be controlled by circuit parameters. It is achieved in an APT circuit, which could be experimentally realized by a series of RLC resonators connected by resistive coupling. The system can achieve more precise control based on the circuit platform, compared to the optics. Our theoretical results exploit calculations based on CMT, Kirchhoff laws and experimental result on a PCB. To achieve the phase difference between the two excitations, we have designed phase-shifting circuits, which can modulate the acquired phase difference virtually continuously from to 0. At the same time, we also adjust the distance between two excitations and observe the effect of the phase difference and distance on the discrete interference. Our work shows the Young's double-slit phenomenon in electrical circuits, and broaden the research perspective of interference.

**Acknowledgements**

The authors thank for the support by NUPTSF (Grants No. NY220119, NY221055).

# APPENDIX A: THE CURCUIT STRUCTURE AND WORKING PRINCIPLE OF THE PHASE-SHIFTING CIRCUIT

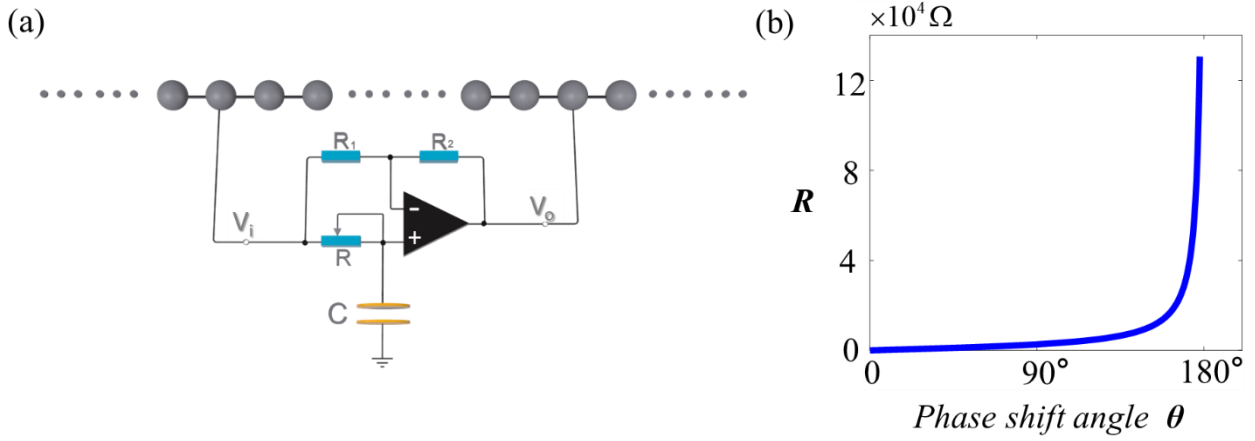

Fig. S1. (a) The structure of phase-shifting circuit; (b) The relationship between variable resistance value and phase shift angle (the phase-shifting effect)

In order to observe the effect of the phase difference between the two excitations on the interference, we designed a phase-shifting circuit with the structure as shown in Fig. S1(a). The ball represents the resonator. The input and output terminals of the phase-shifting circuit are respectively connected to two nodes in the circuit chain that need to have excitation. The initial voltage is added to the capacitance of the resonator at the input end, then through the phase shift circuit, the output end has the initial voltage of the set phase lag. The specific principle is as follows.

From the circuit, we can get $V_+ = \dfrac{1}{1+j\omega RC}V_i$; $V_- = \dfrac{R_2}{R_1+R_2}V_i + \dfrac{R_1}{R_1+R_2}V_o$. According to $V_+ = V_-$, while $R_1 = R_2 = R_0$, we obtain the transfer function

$$\frac{V_o}{V_i} = T(j\omega) = \frac{1-j\omega RC}{1+j\omega RC} = \frac{1/RC - j\omega}{1/RC + j\omega}. \tag{S1}$$

Among Eq. (S1), the denominator is $\dfrac{1}{RC} - j\omega = m_1\angle\phi_1$, while the numerator is $\dfrac{1}{RC} + j\omega = M_1\angle\theta_1$.

Since $m_1 = M_1$, it is clear that the magnitude $T(j\omega) = \dfrac{m_1}{M_1} = 1$ for all values of frequency. The designed phase angle is

$$\theta_d = \theta_1 - \phi_1 = \arctan\left(\frac{\omega}{-1/RC}\right) - \arctan\left(\frac{\omega}{+1/RC}\right). \tag{S2}$$

The range of the two angles will be $-90° < \theta_1 < 0°$ and $0° < \phi_1 < 90°$. Since $\theta_d = \theta_1 - \phi_1$, the range of $\theta_d$ will be $-180° < \theta_d < 0°$.

Therefore, we can obtain the amplitude frequency and phase frequency curve of the phase shift circuit, through the method of which the signal amplitude remains unchanged, and the phase realizes $0°$ to $180°$ adjustable lag. Here, we need to realize the phase difference at any angle by changing the resistance $R$ to the resonance frequency. The corresponding relationship between $\theta_d$ and $R$ is shown in Fig. S1(b). When performing theoretical calculations based on CMT, we need to set one of the initial values as 1 and the other as a complex number to indicate the phase difference of the excitations, with a modulus of 1 and an angle equal to the desired angle.

# APPENDIX B: SUPPLEMENT TO EFFECTS OF PHASE DIFFERENCE AND DISTANCE ON INTERFERENCE BY THE RESULTS OF MATLAB SIMULIINK

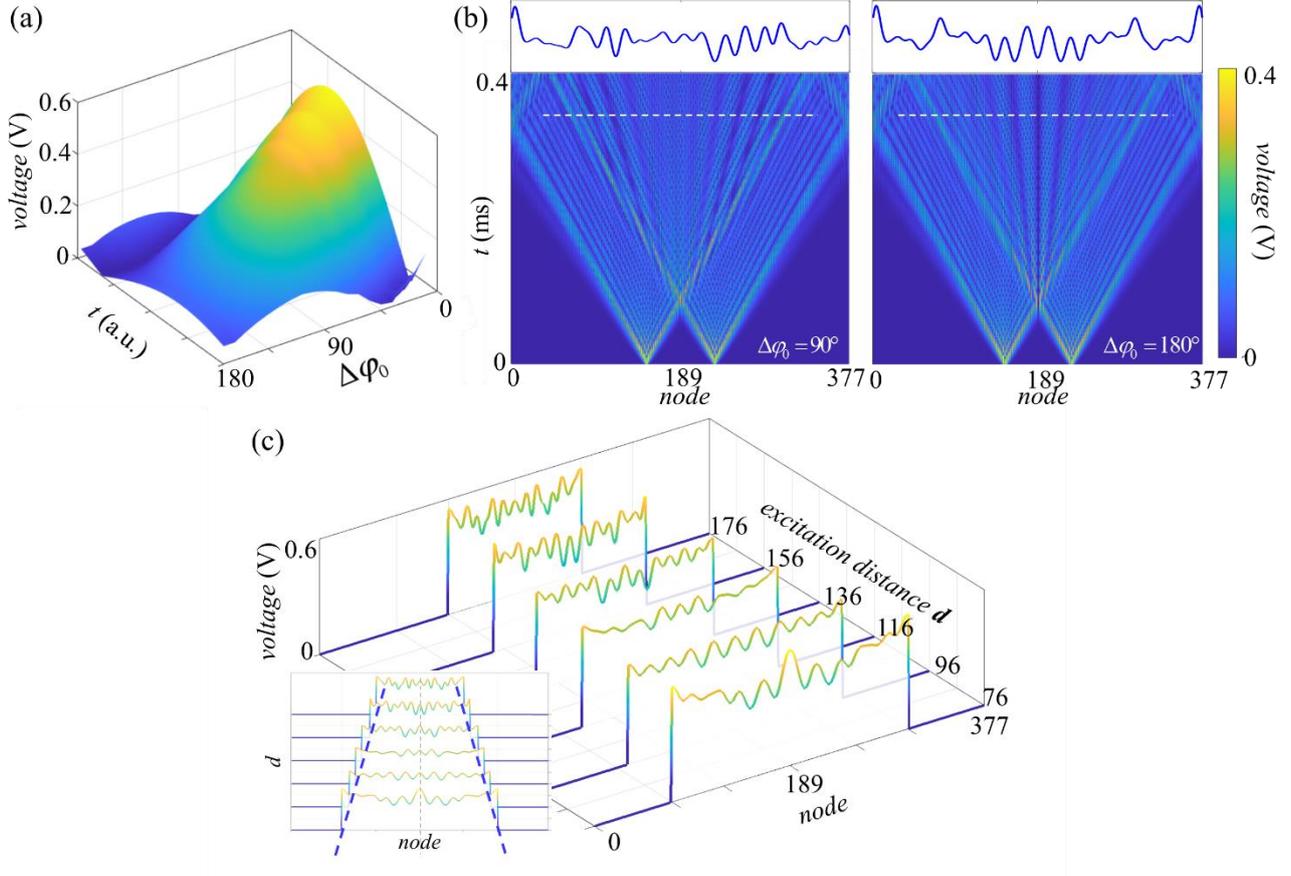

Fig. S2. Effects of phase difference and distance between two excitations on interference by the results of MATLAB Simulink. (a) Influence of different phase differences between two excitations on the voltage near the interference point. (b) Voltage distribution for excitation phase differences of 90 and 180 degrees, respectively. (c) Influence of different distances between two excitations on the voltage at various nodes at a certain moment.

The results from the MATLAB Simulink are presented in Fig. S2, largely consistent with the results of the CMT presented in the main text Fig. 4. The voltage change near the point *M* in Fig. 3 is shown in Fig. S2(a). As the phase difference between the two excitation increases, the voltage changes from interference strengthening to interference weakening, and the voltage evolution at $\Delta\varphi_0 = 90°$ and $\Delta\varphi_0 = 180°$ is shown in Fig. S2(b). When the two excitation distance *d* increases, the region of interference decreases and the stripe spacing decreases, shown in Fig. S2(c).

# APPENDIX C: CIRCUIT MEASUREMENT AND TESTING INSTRUMENT

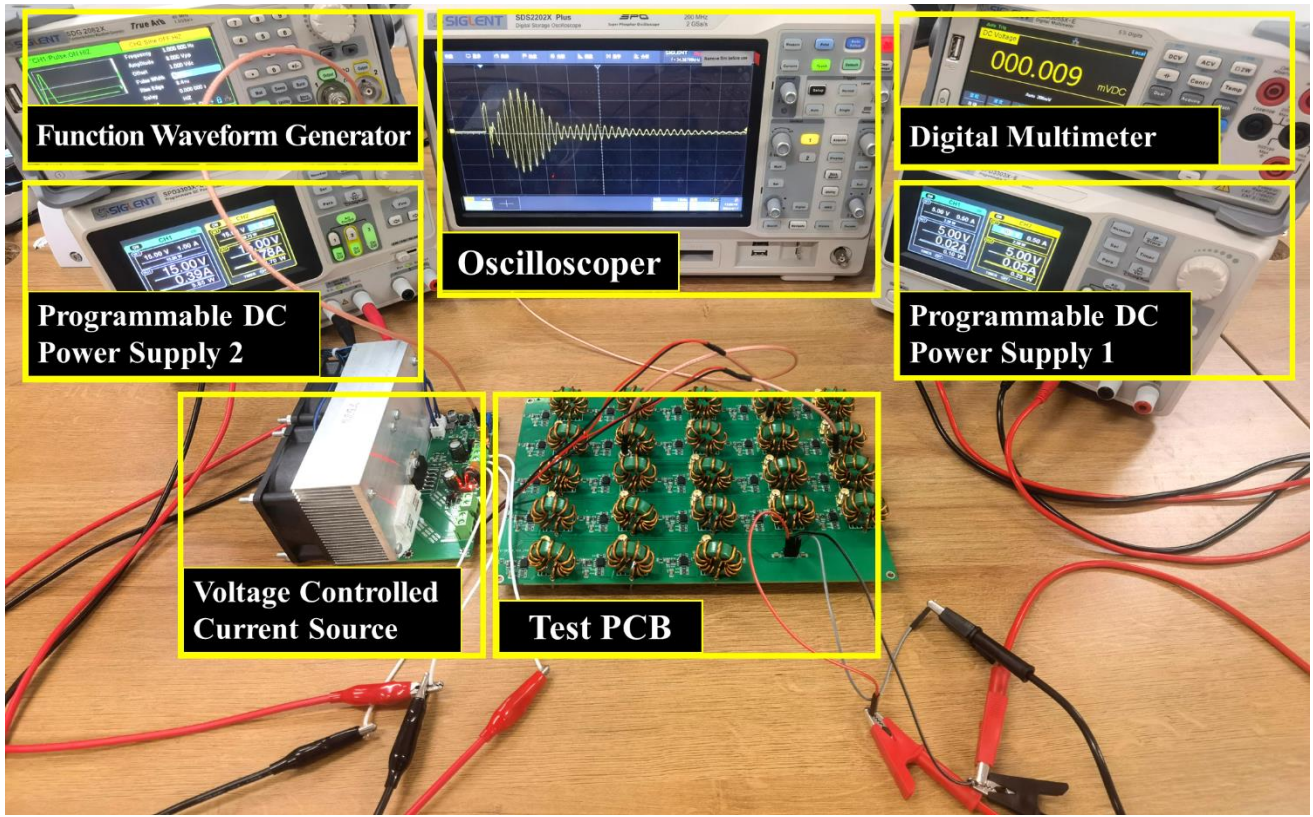

FIG. S3 Photos of PCB and testing instruments.

We use JLC EDA software for PCB development, creating designs for PCB composition, stacking layout, inner layers, and grounding to prevent signal crosstalk during experimental testing. We utilize SMA interfaces and coaxial cables to connect the oscilloscope and signal generator without interference. Our laboratory is equipped with a complete set of SIGLENT instruments for measurement including the SDS2202X Plus Oscilloscope, SDG 2082X Function Waveform Generator, SPD3303X-E Programmable DC Power Supply, and SDM3055X-E Digital Multimeter.